\documentclass[12pt]{article}
\usepackage[bookmarks=false,colorlinks=true,linkcolor=blue,citecolor=blue]{hyperref}
\usepackage{natbib}
\usepackage{amsmath,mathrsfs}
\usepackage{ulem}
\usepackage{color}
\usepackage{graphicx}
\usepackage{multirow}
\usepackage{algorithm,algorithmic}
\usepackage{setspace}
\makeatletter
\newcommand{\vast}{\bBigg@{4}}

\newcommand{\rmn}[1]{{\mathrm{#1}}}
\usepackage{caption}
\newcommand{\footremember}[2]{%
   \footnote{#2}
    \newcounter{#1}
    \setcounter{#1}{\value{footnote}}%
}
\newcommand{\footrecall}[1]{%
    \footnotemark[\value{#1}]%
} 

\begin{document}

\title{Semi-parametric Bayes Regression with Network Valued Covariates}
\author{%
    Xin Ma\footremember{bios}{Department of Biostatistics and Bioinfomatics, Emory University}%
    \and Suprateek Kundu\footrecall{bios}%
    \and Jennifer Stevens\footnote{Department of Psychiatry and Behavioral Sciences, Emory University}%
}
\date{}
\maketitle

\begin{abstract}
There is an increasing recognition of the role of brain networks as neuroimaging biomarkers in mental health and psychiatric studies. Our focus is posttraumatic stress disorder (PTSD), where the brain network interacts with environmental exposures in complex ways to drive the disease progression. Existing linear models seeking to characterize the relation between the clinical phenotype and the entire edge set in the brain network may be overly simplistic and often involve inflated number of parameters leading to computational burden and inaccurate estimation. In one of the first such efforts, we develop a novel two stage Bayesian framework to find a node-specific lower dimensional representation for the network using a latent scale approach in the first stage, and then use a flexible Gaussian process regression framework for prediction involving the latent scales and other supplementary covariates in the second stage. The proposed approach relaxes linearity assumptions, addresses the curse of dimensionality and is scalable to high dimensional networks while maintaining interpretability at the node level of the network. Extensive simulations and results from our motivating PTSD application show a distinct advantage of the proposed approach over competing linear and non-linear approaches in terms of prediction and coverage.
\end{abstract}
\noindent%

\section{Introduction}
\label{s:intro}
During their lifetime, 60.7\% of men and 51.2\% of women experience at least one potentially traumatic event \citep{kessler1995posttraumatic}. Of those experiencing potentially traumatic events, 10-40\% develop psychiatric symptoms of clinical relevance \citep{breslau2004trauma,breslau2009epidemiology} such as post-traumatic stress disorder (PTSD). PTSD is one of the most common mental disorders in the USA and results in significant impairments of psychological and physical health \citep{kessler1995posttraumatic}. Previous studies have implicated several PTSD-related brain areas including amygdala, hippocampus, medial prefrontal cortex (mPFC), anterior cingulate cortex (ACC) and insula \citep{brown2014altered}, which are associated with emotion, memory and executive functions \citep{shalev2017post}. However, it is increasingly recognized that these areas do not act in isolation, and that the disease severity can often be better accounted for by considering the interactions between pairs of brain regions. Such interactions or co-activations can be captured via functional connectivity (FC) that encodes the temporal coherence between regions \citep{smith2011network}. The brain regions and their functional connections form what we call the brain network.

\begin{figure}
  \begin{tabular}{cc}
  \centering
    \includegraphics[width=0.35\linewidth]{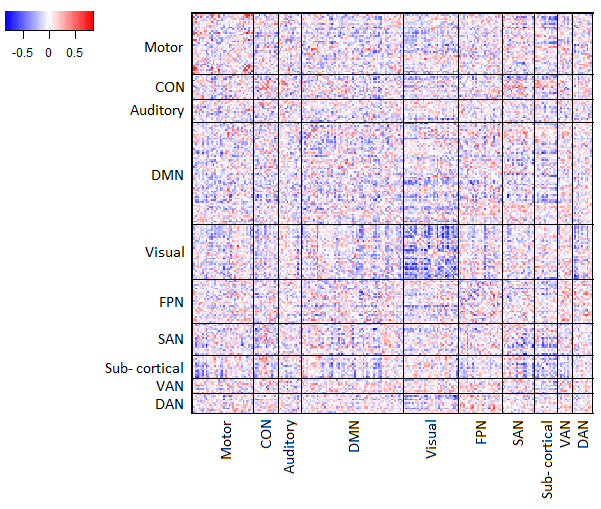} & \includegraphics[width=0.35\textwidth]{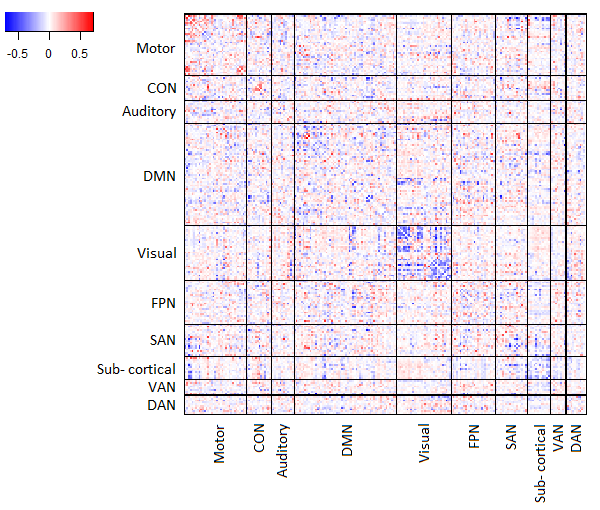} \\
  \end{tabular}
    \caption{Differences in absolute correlations (left panel) and fitted edge probabilities (right panel) between participants with highest and lowest resilience score. Fitted edge probabilities can be viewed as a denoised version of the absolute correlations. Visual, salience and sub-cortical functional networks see high proportions of edges with large difference.}
    \label{fig:abscor}
\end{figure}

Hence, there is a strong underlying justification to model PTSD outcomes in terms of brain network. As an example, PTSD based network differences have been discovered in terms of the salience network \citep{sripada2012neural} and dorsal attention network \citep{hayes2009alterations}. Moreover, in our motivating Grady Trauma project (GTP) application, we discovered clear differences in brain connectivity between individuals with high and low PTSD resilience (see Figure \ref{fig:abscor}). Although there has been some progress in classifying the disease status in PTSD studies based on neuroimaging biomarkers \citep{fenster2018brain}, such existing approaches may not be fully satisfactory given the fact that the definition of PTSD itself has been a stumbling block in psychiatry. Hence instead of focusing on classification, it is appealing to develop prediction approaches for continuous clinical phenotypes in PTSD based on the brain network and other risk factors. However, such network valued regression is not straightforward due to the high-dimensional and complex nature of the brain networks, the unknown interactions between the brain network and environmental exposures such as trauma, as well as the possibly highly non-linear relationship between the clinical phenotype and the network. Supplementary factors such as own/family history of psychiatric disorders, the experience of a traumatic event, and female gender were shown to be associated with an elevated PTSD risk \citep{kessler1995posttraumatic} and hence need to be accounted for.

A common approach to modeling outcome based on network valued covariates is to use summary network measures as covariates (see, for example, \cite{bullmore2009complex}  and  references  therein). However, the success of such an approach depends heavily on the choice of the network metrics. Moreover, these approaches have reduced exploratory value and possibly less accuracy due to a decreased resolution of the summary statistics. Another alternative is to include all the edges in the network as a vectorized predictor, and use these high dimensional features for modeling the clinical phenotype \citep{richiardi2011decoding,craddock2009disease}.  Although penalized regression approaches \citep{tibshirani1996regression} and Bayesian shrinkage \citep{park2008bayesian, chang2018scalable} may be used to model the regression coefficients in these high-dimensional applications, these approaches treat the edges as interchangeable and fail to respect the inherent structure of the network that may show properties such as small-worldedness \citep{bassett2006small} or other patterns of organization. Disregarding these inherent structures and the associated correlations may lead to sub-optimal performance.   Recent developments such as \cite{guha2018bayesian} address this issue by using a tensor-based representation for the regression parameters corresponding to the edges in the network. However, their approach still requires sampling as many regression coefficients as there are edges, and hence is challenging to implement for high dimensional networks including hundreds of regions containing tens of thousands of edges (as in the Power network \citep{power2011functional} having 264 regions). 

More importantly, existing approaches such as the ones referred to above, model the outcome as a function of a linear combination of the edges. While linearity assumption may be valid for some scenarios, and enables efficient computation and interpretability, they may not be applicable to more general models that input edge strengths as explanatory variables and to more diverse settings in general. Linear models also cannot accommodate unknown interactions between the brain network and environmental exposures such as childhood and adult trauma. Accounting for such interactions is crucial according to neurobiological models that conceptualize the symptoms of PTSD as correlates of dysfunctional stress reaction to traumatic events \citep{heim2009neurobiology}, where neural cues become associated with the traumatic event that may trigger a conditioned fear response, and failure to extinguish such a response is thought to lead to the persistence of symptoms \citep{vanelzakker2014pavlov}. In our motivating GTP application, we have seen non-linear associations between PTSD resilience (a clinical measure determining the susceptibility to PTSD after childhood trauma) and some edge strengths represented by pairwise correlations. Thus it is imperative to develop flexible predictive approaches that enable non-linear association between the clinical phenotype and the high-dimensional brain network along with other risk factors. One alternative is to directly use a Gaussian process regression based on the entire edge set. However due to the high-dimensionality of the network, it may be challenging to sample a large number of lengthscale parameters under an anisotropic Gaussian process model via  Metropolis-Hastings updates \citep{bhattacharya2014anisotropic}.

An alternative approach to tackle the problem is through non-linear manifold regression framework that first projects the high-dimensional feature space on a lower dimensional manifold, and subsequently uses flexible regression approaches to predict the outcome based on the latent variables. Frequentist examples of reducing the dimension of feature space include principal components analysis and more elaborate methods that accommodate non-linear subspaces, such as isomap \citep{tenenbaum2000global} and Laplacian eigenmaps \citep{belkin2003laplacian,guerrero2011laplacian}. Bayesian manifold approaches characterizing predictive uncertainty have also been developed. \cite{page2013classification} proposed a Bayesian nonparametric model for learning of an affine subspace in classification problems. More flexible Bayesian methods that accommodate non-linear subspaces include Gaussian process latent variable models (GP-LVMs) \citep{lawrence2005probabilistic,kundu2014latent}. However, there may be a heavy computational price for learning the number and distribution of the latent variables, and for learning the mapping functions while keeping identifiability restrictions. These factors typically restrict these approaches to a small number of features. \cite{snelson2012variable} also considered manifold regression for big data, comprising feature vectors via pre-multiplying with a short and fat projection matrix. These approaches as well as other manifold regression methods (see \cite{bickel2007local}; \cite{aswani2011regression}) often lack scalability even for moderate number of features. \cite{yang2016bayesian} developed a scalable approach that relied on usual Gaussian process regression without attempting to learn the mapping  to  the  lower-dimensional  subspace. \cite{guhaniyogi2016compressed} extended this approach to more flexible settings. However, these existing methods are not designed for estimating manifold for high dimensional networks which is our focus, thus cannot be directly applied to our problem of interest. Moreover, the lower dimensional projected features constructed under existing manifold based methods are typically not interpretable in terms of the the original features, which may be restrictive in neuroimaging applications where it is often important to identify important brain regions that drive a mental illness.

In order to close these gaps, we develop a two-stage semi-parametric Bayesian manifold regression approach for high dimensional network-valued covariates, where we first project the high-dimensional brain network on a lower dimensional manifold, and then use these projected features to predict the clinical phenotype via a flexible Gaussian process regression. The lower dimensional manifold corresponding to the brain network is obtained under the latent scale model \citep{hoff2005bilinear}. This model projects each node in the network into a lower dimensional Euclidean space in a way that the inner product between the projected nodes is representative of the corresponding edge probability in the network. Thus the lower dimensional manifold has one dimension corresponding to the number of nodes in the network ensuring the interpretability at the node level. We denote our approach as latent scale Gaussian process regression (ls-GPR). It allows for quantification of uncertainty given the latent scales representing the network, and also enables non-linear associations with the outcome, as well as unknown interactions between the network and other exposure variables. We develop an efficient optimization algorithm for estimating the latent scales from the given network data in the first stage, by leveraging the Polya-Gamma data augmentation scheme proposed in \cite{polson2013bayesian}. The Gaussian process regression in the second step is implemented via standard tools in literature.

We choose a two stage model over a joint model that learns the latent scales as well as the relationship of the outcome with the latent scale simultaneously, based on computational considerations. In particular, it is challenging to derive a closed form for the posterior distribution of the latent scales under a joint model, especially when the number of nodes is large. One can discretize the latent scales to facilitate computational updates, or alternatively use Metropolis-Hastings based strategies or their more efficient variants \citep{robert2015metropolis}. However, both strategies have drawbacks in high dimensions: the former may lead to shrinkage of prior support, while the latter may result in inefficient mixing. In order to overcome the above difficulties, we propose the two-step approach that results in superior performance in our motivating PTSD application, as well as extensive simulations. In particular, the proposed approach has superior prediction and coverage in test samples.

\section{Methods}
\begin{figure}
    \centering
    \includegraphics[width=0.5\textwidth]{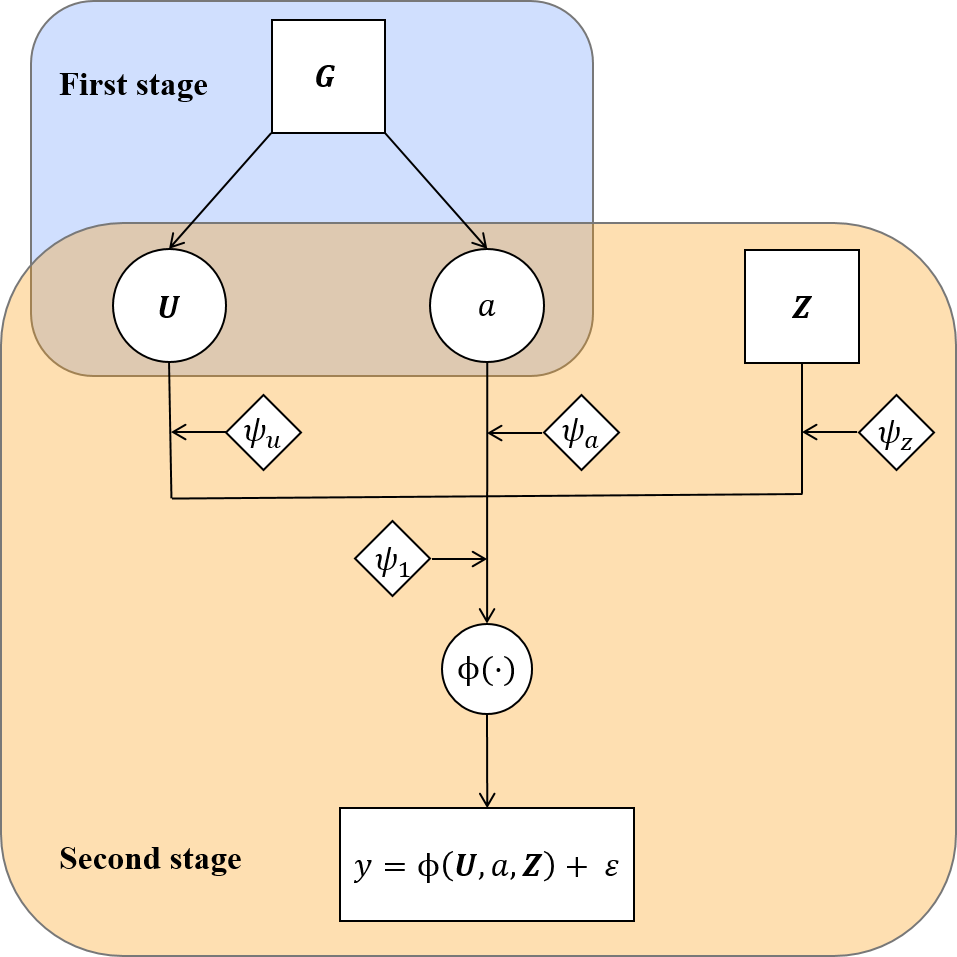}
    \caption{Schematic Diagram of the Two-stage Model}
    \label{fig:schematic}
\end{figure}

We have data on $n$ participants. For the $i$th participant ($i=1,\ldots,n$), the data includes a continuous scalar variable $y_i\in \Re$ representing the clinical outcome of interest, an undirected brain network having $p$ regions of interest (ROIs) represented as a symmetric binary matrix $G_i(p\times p)$, with $g_i(k,l)=1/0$ depending on whether the edge $(k,l), k\ne l,$ is present or absent in the network, and supplemental covariates $\boldsymbol{z}_i$ representing environmental exposures and demographic factors. The binary network $G$ can be based on structural or functional connectivity, while the number of regions depends on the chosen atlas ($p=264$ under the Power atlas for our applications). We denote the vector of elements in the upper triangle excluding the diagonals for $G_i$, or edge set, as ${\bf e}_i$ of length $p(p-1)/2$. The diagonal elements are excluded since they do not represent connections between distinct nodes and are irrelevant to our problem of interest. The method is described in detail below, and Figure \ref{fig:schematic} provides a diagrammatic illustration of our two-stage model.

\subsection{Model formulation}
\underline{\noindent\textit{First Stage: Latent Scale Representation of Brain Networks:}}
Our goal is to have a parsimonious probability model for the binary networks represented by the edge sets ${\bf e}_i, i=1,\ldots,n$. Clearly there are $2^{p(p-1)/2}$ possible values and hence the parameter space grows exponentially with the number of nodes. In order to tackle this problem, we seek to project these parameters into a lower dimensional space via a meaningful mapping that avoids restrictive assumptions and fits the data reasonably well. Motivated by the above considerations, we represent the edge probabilities in terms of node level latent scales. In particular,
\begin{equation}
P\big({\bf e}_i\big) = \prod_{k<l,k,l=1}^p \pi_{i,kl}^{e_{i,kl}}(1-\pi_{i,kl})^{1-e_{i,kl}}, \mbox{ } \log(\frac{\pi_{i,kl}}{1-\pi_{i,kl}}) = a_{i} + {\bf u}_{ik} \Lambda_i {\bf u}_{il}^T, \mbox{ } i=1,\ldots,n,
\label{eqt:stage1}
\end{equation}
where $a_i$ denotes the subject-specific intercept common across edges, ${\bf u}_{ik}=(u_{ik,1},\cdots,u_{ik,d})$ is the vector of latent scale for node $k$ of the $i$th participant of length $d$, and $\Lambda_i$ represents the $d\times d$ diagonal weight matrix having all positive elements, and which controls the contribution of the latent scales to the inner product. The intercept term controls the overall density of the network and is learnt by pooling data across all edges. The latent scale ${\bf u}_l$ captures the importance of node $l$ in the network. If both nodes $l$ and $k$ have activations in the same directions, captured via $u_{l,d'}$ and $u_{k,d'}$ having the same signs, then they can be construed as functionally connected. We note that the latent scales are only identifiable up to a rotation in $\Re^d$. In other words, by rotating the latent scales the inner products (and hence the edge probabilities) in model (\ref{eqt:stage1}) stay unchanged since ${\bf u}_{ik}\Lambda_i{\bf u}_{il}^T = {\bf u}_{ik}Q\Lambda_i({\bf u}_{il}Q)^T$ for any orthonormal matrix $Q$ and diagonal $\Lambda_i$. Thus in order to preserve identifiability with respect to rotation, we fix the first element of all latent scales (0.5 in our simulations and real data analysis).

Model (\ref{eqt:stage1}) results in a dramatic reduction in the number of parameters from $p(p-1)/2$ to $(p\times d+1)$. In particular, the latent scales $U_i=({\bf u}_{i1}^T,\ldots,{\bf u}_{ip}^T)^T,$ for participant $i$, have dimension $p\times d$ where $p$ is the number of brain regions and $d$ is the dimension of the latent scale or the intrinsic dimension that needs to be determined. In general, finding the intrinsic dimension of the manifold is a difficult problem \citep{yang2016bayesian}. One approach is to treat $d$ as a tuning parameter and determine the optimal dimension by choosing the one with smallest prediction mean squared error from cross validation of the following second stage regression. We also illustrate via simulations that the prediction accuracy is pretty robust to a reasonable range of values for $d$. 

Model (\ref{eqt:stage1}) adapts the dot product characterizations of edge probabilities for a single network \citep{hoff2005bilinear,hoff2008modeling} to the case of multiple networks. In latent space modeling of a single network, this representation has been shown to provide a more general characterization of interconnection structures and network properties than stochastic block model \citep{nowicki2001estimation} and latent distance model \citep{hoff2002latent}. Similar model for undirected binary network has appeared in \cite{durante2017nonparametric}, who used a mixture of latent scale probabilities to model a population of networks. For our PTSD application, it may be too simplistic to assume that groups of participants in the sample 
share the same network exactly, due to the well-known inherent heterogeneity in PTSD \citep{lanius2006review}. Moreover, our primary objective is to predict the clinical outcome based on the network, which requires a distinct lower dimensional representation for each network. Hence for our application, having a separate model for each participant's network seems meaningful, although one could potentially use a subset of shared parameters across participants to learn common patterns in the network (see \citet{lukemire2017bayesian} for some context on joint learning of multiple networks). 

\underline{\textit{Second Stage: Latent Scale Gaussian Process Regression:}}
In this stage, we propose to use a flexible Gaussian process regression framework for the continuous scalar response based on the lower dimensional node-specific representation of network from the first stage model and other covariates as follows
\begin{equation}
y_i = \phi(U_i, a_i, \boldsymbol{z}_i) + \epsilon_i, \mbox{ } \epsilon_i \overset{\mathrm{iid}}{\sim}  \mathrm{N}(0, \tau^{-1}), \mbox{ } i=1,\cdots,n.
\label{eqt:stage2}
\end{equation}
where $\epsilon_i$ denotes the residual error normally distributed with precision $\tau\sim \pi(\tau)$, $\phi(\cdot)$ denotes the unknown mean that is a function of the brain network via the latent scales $U_i$ and the intercept term $a_i$ in model (\ref{eqt:stage1}), as well as supplementary demographics and environmental exposures ${\boldsymbol z}_i$. The function $\phi(\cdot)$ is assumed to have a Gaussian process prior with mean $\mathbf{0}$ and covariance kernel $K$ that has the following structure:
\begin{equation}
K(i,i^\prime) = \psi_1 exp\Big\{-\psi_u \big|\big|U_i - U_{i^\prime}\big|\big|_F^2 - \psi_a (a_i - a_{i^\prime})^2 - \psi_z \big|\big|\boldsymbol{z}_i - \boldsymbol{z}_{i^\prime}\big|\big|_2^2\Big\}, \mbox{ } i\ne i'
\label{eqt:kernel}
\end{equation}
where $||\cdot||_F$ and $||\cdot||_2$ denote the Frobenius and $L_2$ norms respectively, $\psi_1$ denotes the scale parameter controlling the variance of the mean function, and $\psi_u, \psi_a$ and $\psi_z$ respectively denote the distinct lengthscale parameters corresponding to the latent scales, the intercept term and the supplementary covariates. The lengthscale parameters control the smoothness of the curve. By specifying three distinct parameters, we take into account that the covariates may potentially lie on very different scales and influence the smoothness to varying degrees. The Gaussian process prior for the mean function allows flexible non-linear relationships between the outcome and the covariates, and can also accommodate unknown interactions between the covariates.

\subsection{Computation Framework}
The estimation of the first stage model is implemented through an EM algorithm with data augmentation utilizing Theorem 1 in \cite{polson2013bayesian}. The EM algorithm is more computationally efficient compared to the previous Markov chain Monte Carlo (MCMC) implementation \citep{hoff2005bilinear}, and leads to significant speed-ups in our PTSD application involving high-dimensional networks and a moderate number of subjects.

We use an MCMC to estimate the parameters for the second stage regression model, with initialization at their maximum likelihood values. In our experience this leads to faster mixing and improved estimates. The MLE for the parameters ($\tau, \psi_1, \psi_u, \psi_a, \psi_z$) are derived via a gradient descent algorithm with Armijo line search \citep{armijo1966minimization}, using the marginal likelihood    $\boldsymbol{y}\sim\mathrm{N_n}(\mathbf{0}, K + \tau^{-1}\mathrm{I}_n)$. With properly assigned priors, we can then develop a Gibbs sampler to obtain the posterior distributions for these parameters. We assign a conjugate Gamma prior with parameters $a_\tau$ and $b_\tau$ to the noise precision $\tau$. We assign an inverse Gamma prior with parameters $a_{\psi_1}$ and $b_{\psi_1}$ to $\psi_1$, also conjugate. For the lengthscale parameters $\psi_u$, $\psi_a$ and $\psi_z$, we need to have Metropolis-Hastings steps within the Gibbs sampler.

The prediction for new test samples can be done by first estimating the latent scales with the EM algorithm, and subsequently feeding them into the fitted Gaussian process regression model. The details of the computation framework can be found in the Appendices.

\section{Simulation Studies}
In this section, we tested the prediction performance of our model in different scenarios including both simulated binary networks and those obtained from real data. When fitting first stage model, we fixed the dimension of the latent scales to $d=10$ which provided good results under the regression model for a variety of scenarios. 

\subsection{Data Generation}
\underline{\textit{Scenario 1:}}
In this part of simulation, the edge set $\boldsymbol{e}_i$ was generated according to the first stage model (\ref{eqt:stage1}). In particular, the intercept term was simulated as $a_i\sim\rmn{N}(0,2^2)$, and elements of the latent scales $U_i (p\times d_0)$ were generated independently from $\rmn{N}(0,1^2)$, with $d_0$ taking various values. $\Lambda_i$ was fixed at $\rmn{I}_{d_0}$. The response $y_i$ was generated as $y_i = \sum_{k^* < l^*, (k^*,l^*)\in \mathcal{C}} \sin\big[\pi_{i,k^*l^*}\eta_{i,k^*l^*}\big] + \epsilon^*_i, \epsilon^*_i\sim \mathrm{N}(0,0.5^2),\eta_{i,k^*l^*}\sim N(2,1)$ where $\mathcal{C}$ denoted the set containing edges between $100S_p\%$ nodes, $S_p\in(0,1)$ was the sparsity level for generating the response, and $\eta_{i,k^*l^*}$ was a random perturbation term multiplied to the edge probabilities. This scheme was repeated to generate binary network and response for 50 training and 50 testing samples. Three levels of $S_p$ were chosen including low (0.25), medium (0.5) and high (0.75). The responses were standardized with the mean and standard deviation of the training samples.

\noindent\underline{\textit{Scenario 2:}}
Here, we generated the binary network based on resting state fMRI connectivity matrix of the Grady Trauma Project. A description of the study can be found in Section \ref{subs:GTP}.% There were a total of 81 participants with resting state fMRI time series preprocessed into Power's system with 264 ROIs \citep{power2011functional}.
We calculated the resting state network for each subject using the graphical lasso algorithm  \citep{friedman2008sparse} under varying sparsity levels corresponding to regularization parameter values $\lambda=0.05,0.10,0.15,$ and $0.20$, with a larger $\lambda$ value corresponding to a sparser network. Once the networks were obtained as above, the response was generated with varying percentage of randomly selected ROIs as in Scenario 1. The binary networks and responses were randomly split equally into training (50) and testing (50) sets.

\subsection{Methods to Compare}
We reported results under the proposed approach using MLE based estimates (ls-GPR1) for the Gaussian process regression (GPR), as well as MCMC implementation (ls-GPR2). Our methods were compared to the linear shrinkage methods including lasso \citep{tibshirani1996regression}, ridge \citep{hoerl1970ridge}, elastic net \citep{zou2005regularization}, and Bayesian horseshoe prior \citep{carvalho2010horseshoe}, which all used the full edge set as the predictors. The models were implemented through \texttt{R} packages \texttt{glmnet} \citep{friedman2010regularization} and \texttt{monomvn} \citep{monomvn}. We also compared with non-linear approaches that used GPR on the full edge set (edge-GPR), on the reduced representation from principal component analysis (pca-GPR), and on the reduced representation from Laplacian Eigenmap \citep{belkin2003laplacian} (mf-GPR). The dimension reduction of the pca-GPR and mf-GPR methods were implemented using \texttt{R} function \texttt{prcomp} and \texttt{R} package \texttt{dimRed} \citep{dimRed} respectively. We applied the squared exponential kernel \citep{williams2006gaussian} to all GPR approaches. We evaluated prediction performance in terms of predictive MSE on the test samples for all approaches, as well as coverage and the interval width for the test samples under ls-GPR2 and the competing GPR approaches. Here the coverage was defined as the ratio of the number of intervals covering the true value and the total number of test samples.  

\subsection{Results}
\begin{figure}
    \centering
    \includegraphics[width=0.85\textwidth]{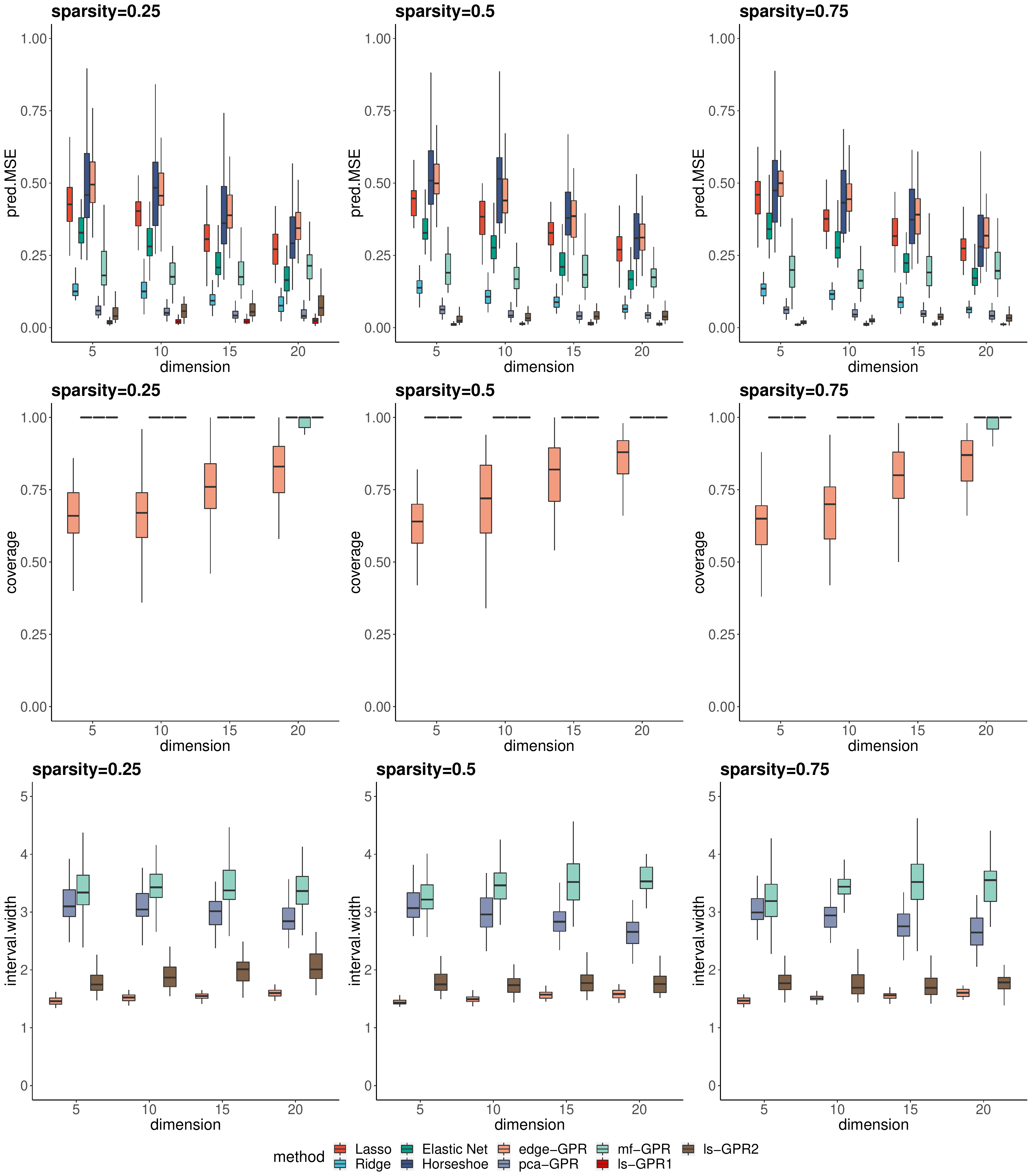}
    \caption{Boxplots for MSE, coverage and interval width of simulation scenario 1. The horizontal axis represents the true dimension of the latent scales, while the estimated dimension is kept at 10. Non-linear methods generally perform better than linear competitors, except for edge-GPR. The proposed ls-GPR methods have the overall best performance. The pca-GPR sometimes has comparable predictive MSE and good coverage, but with much wider intervals.}
    \label{fig:sce1}
\end{figure}

The prediction and coverage results for scenario 1 are presented in Figure \ref{fig:sce1}. The horizontal axis represents the true dimension of the generated latent scales $d_0$. As the true dimension increases, the dimension of the estimated latent scales is kept at $d=10$ for model fitting purposes. The performance of our methods maintains consistency even when the true dimension exceeds that of the estimated latent scales. This provides evidence for the robustness of our prediction performance. We observe that the linear regression approaches have poor prediction performance compared to the non-linear competitors except for edge-GPR which also perform poorly. This is not surprising given that the true response has a non-linear relationship with the edge set. Our ls-GPR1 method always has the lowest predictive MSE compared to other approaches, while our ls-GPR2 method leads to a superior predictive MSE over competing methods when the sparsity level of the true regression model is 0.5 or 0.75. When the sparsity level is 0.25, our ls-GPR2 method may or may not have a smaller predictive MSE compared to pca-GPR. Both our ls-GPR2 and pca-GPR have a coverage close to one for varying sparsity levels of the true network. However, our ls-GPR2 consistently has narrower intervals, which illustrates a greater certainty when predicting test samples. On the other hand, the edge-GPR and the mf-GPR methods consistently have the least favorable prediction performance among the non-linear approaches. Moreover, the mf-GPR has comparably high coverage which comes at the cost of substantially wider predictive intervals, highlighting the lack of precision under this approach. The coverage under the edge-GPR method is alarmingly low, which illustrates the inadequacy of including the entire edge set for prediction.

\begin{figure}
    \centering
    \includegraphics[width=0.85\textwidth]{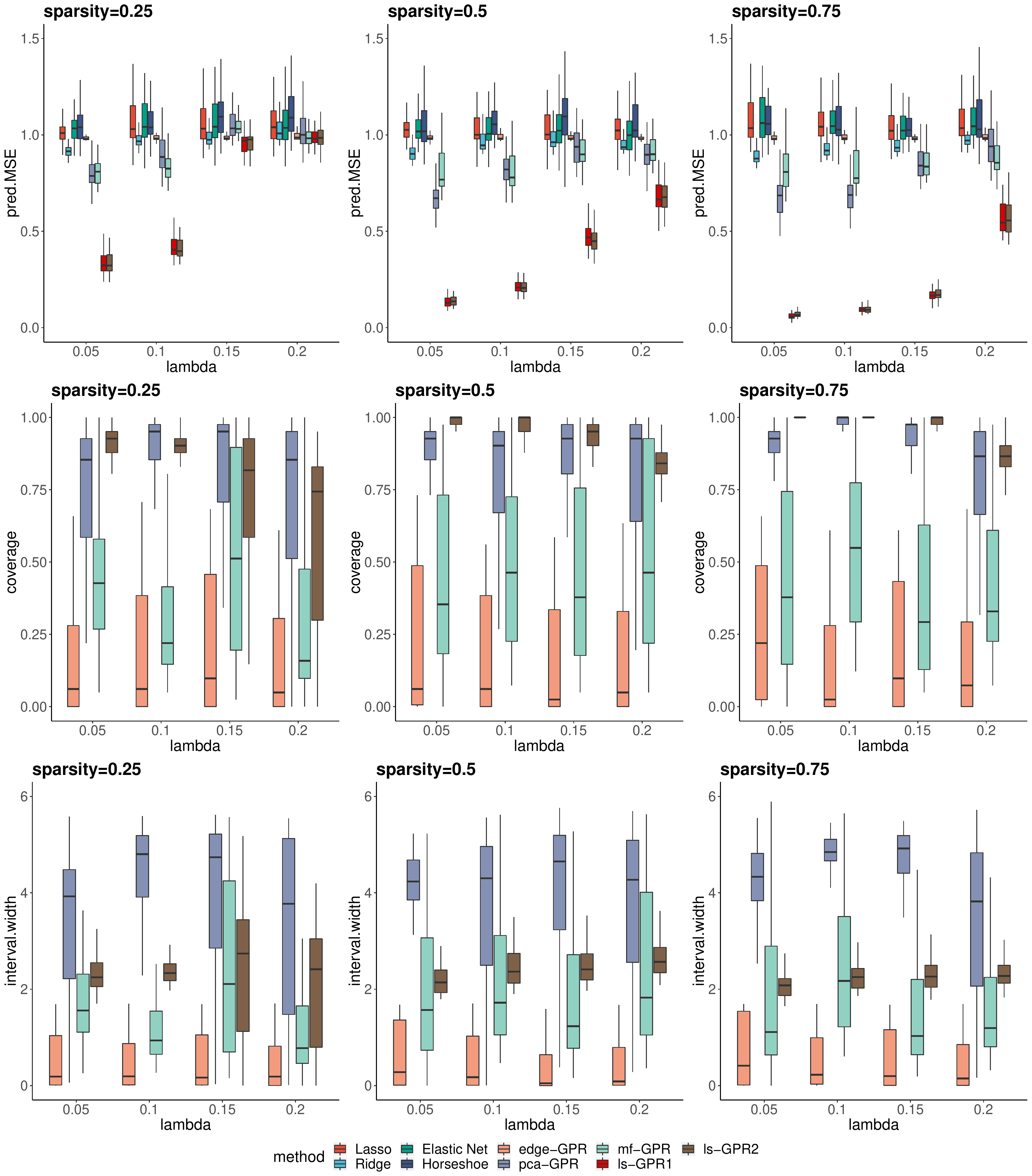}
    \caption{Boxplots for MSE, coverage and interval width of simulation scenario 2. The horizontal axis represents the regularization parameter $\lambda$ controlling density of the binary networks. The proposed ls-GPR methods consistently indicate better prediction performance, especially with smaller $\lambda$. The pca-GPR occasionally has better coverage but at the cost of wider intervals.}
    \label{fig:sce2}
\end{figure}

Figure \ref{fig:sce2} shows the results of scenario 2. The horizontal axis represents the regularization parameter $\lambda$ of the graphical lasso in obtaining the binary networks. Both our ls-GPR1 and ls-GPR2 methods indicate a decisively better prediction performance across all settings. Moreover, our methods perform better as the density of the underlying network increases (smaller $\lambda$ values). The coverage under our ls-GPR2 is sometimes better and in some cases worse compared to the pca-GPR depending on the sparsity levels of the true regression model. However, our ls-GPR2 consistently has a narrower predictive interval compared to the pca-GPR across all settings, highlighting the increased precision of our approach. The edge-GPR and mf-GPR methods have a consistently smaller coverage resulted from very narrow predictive intervals. 

In summary, the above results highlight the predictive advantages using our proposed methods. Combined with the close to optimal coverage and desirable width for the predictive intervals, these results make a clear case for the utility of the proposed approach when modeling complex associations of the clinical outcome with the underlying brain network. 

\section{Real Data Analysis}
\subsection{Grady Trauma Project and Resilience Scores}
\label{subs:GTP}
The Grady Trauma Project (GTP) has recruited African American females to study the risk factors for PTSD in a low-socioeconomic status \citep{stevens2013disrupted,stevens2018episodic}. The resting state functional magnetic resonance imaging (fMRI) data has been obtained and preprocessed following the same protocol as \citet{wang2016efficient}. After removing data with movement or drowsiness issues, we have 81 participants available. For our analysis, we use the whole brain parcellation presented in \cite{power2011functional}, involving 264 ROIs. These regions are further organized into ten functional modules including motor, cingulo-opercular, auditory, default mode, visual, fronto-parietal, salience, sub-cortical, ventral attention and dorsal attention \citep{cole2013multi}. These functional modules have been assigned based on resting state fMRI studies \citep{power2011functional}, which is well-suited for our data.

The GTP study has also acquired data on the Connor-Davidson Resilience Scale \citep{connor2003development} for measuring reselience as individual's ability to thrive in the face of adversity. Our goal is to model resilience as a continuous clinical measure of well-being in PTSD using resting state functional connectivity as well as demographic factors such as the participants' age, and environmental exposure including traumatic events inventory (TEI) score \citep{gillespie2009trauma} and the childhood trauma questionnaire (CTQ) total score \citep{scher2001childhood}. The resilience score of interest is only available for 73 participants, and hence we focus our analysis on this subset. 

As an additional analysis of interest, we also investigate which regions in the brain network contribute to significant differences with respect to resilience. To achieve this, we extend the proposed model in (\ref{eqt:stage2}) by using modified latent scales as $U_i\boldsymbol{\beta} = (\beta_1{\bf u}_{i1}^T,\ldots, \beta_p{\bf u}_{ip}^T)^T$, where $\beta_j\sim \rmn{Bernoulli}(\pi^*), j=1,\ldots,p$. Here $\beta_j$ serves as the indicator for the $j$-th brain region to be included in the regression process. We also specify $\pi^*\sim Be(a_\pi, b_\pi)$, so that the prior inclusion probability for nodes is learnt adaptively from the data. The posterior draws for $\beta's$ can be used to compute posterior inclusion probabilities which indicate the nodes' importance in the network with respect to the outcome. The posterior computation proceeds as in model (\ref{eqt:stage2}) with additional steps included to update $\boldsymbol{\beta}$ and $\pi^*$, as outlined in the Appendix C.

\subsection{Results}
The top panel of Table \ref{tab:CDRISC} shows the prediction performance using resilience score as response. The results are obtained from 100 random splits into training and testing samples. Our proposed methods display lowest predictive MSE across varying network densities represented via different $\lambda$ settings. One sided two sample t-tests show that our proposed methods have significant reduction in MSE at 0.05 level compared to competing methods. The only exceptions are mf-GPR to both ls-GPR1 and ls-GPR2 at $\lambda=0.1$, and mf-GPR to ls-GPR1 at $\lambda=0.15$, but they are still significant at 0.10 level. Our proposed ls-GPR2 has similar or improved coverage as well as narrower predictive intervals compared to mf-GPR, indicating its greater confidence in the predicted values. The proposed ls-GPR2 also has improved coverage compared to pca-GPR across all settings, which is likely due to the narrower intervals under the pca-GPR method. Finally, the edge-GPR method has extremely poor coverage resulting from really tight predictive intervals. 

In summary, although the pca-GPR and the mf-GPR approach occasionally show some advantages in the simulation studies, these methods have less than optimal predictive performance and coverage in the GTP study application. On the other hand, the edge-GPR approach involving the full edge set performs worst among the non-linear regression methods, although the performance is improved from the competing linear regression methods. These results highlight the advantages of non-linear dimension reduction via latent scales under the proposed approach, and project a strong case for implementing the proposed method as a powerful prediction tool in brain network based studies.

\begin{table}
\centering
\resizebox{\textwidth}{!}{
\begin{tabular}{|l|l|ccccccccc|}
\hline
                                                 &                & \textit{Lasso} & \textit{Ridge} & \textit{Elastic net} & \textit{Horseshoe} & \textit{edge-GPR} & \textit{pca-GPR} & \textit{mf-GPR} & \textit{ls-GPR1} & \textit{ls-GPR2} \\
\hline
\multirow{3}{*}{$\lambda=0.05$}                  & \textit{MSE}            & 1.072 & 1.017 & 1.069       & 1.170              & 1.002    & 1.100   & 0.996  &                                                        \textbf{0.934}   & \textbf{0.938}   \\
                                                 & \textit{Coverage}       &     &    &           &                  & 0.138    & 0.801   & 0.810  &       & 0.832   \\
                                                 & \textit{Width} &     &     &           &                  & 0.460    & 4.084   & 4.153  &       & 3.412   \\
\hline
\multirow{3}{*}{$\lambda=0.10$}                     & \textit{MSE}            & 1.085 & 1.034 & 1.105       & 1.193              & 1.010    & 1.075   & 0.969  &                                                    \textbf{0.939}   & \textbf{0.943}   \\
                                                 & \textit{Coverage}       &     &     &           &                  & 0.149    & 0.720   & 0.813  &       & 0.822   \\
                                                 & \textit{Width} &     &     &           &                  & 0.504    & 3.537   & 4.071  &       & 3.314   \\
\hline
\multirow{3}{*}{$\lambda=0.15$}                     & \textit{MSE}            & 1.075 & 1.042 & 1.088       & 1.452              & 1.008    & 1.065   & 0.984  &                                                     \textbf{0.945}   & \textbf{0.935}   \\
                                                 & \textit{Coverage}       &     &     &           &                  & 0.148    & 0.702   & 0.846  &       & 0.851   \\
                                                 & \textit{Width} &     &     &           &                  & 0.468    & 3.429   & 4.313  &       & 3.348   \\
\hline
\multirow{3}{*}{$\lambda=0.20$}                     & \textit{MSE}            & 1.048 & 1.039 & 1.052       & 1.418              & 1.011    & 1.060   & 0.957  &                                                     \textbf{0.925}   & \textbf{0.919}   \\
                                                 & \textit{Coverage}       &     &     &           &                  & 0.151    & 0.588   & 0.802  &       & 0.799   \\
                                                 & \textit{Width} &     &     &           &                  & 0.492    & 2.885   & 4.054  &       & 3.150 \\
\hline
\end{tabular}
}

\bigskip

\resizebox{0.9\textwidth}{!}{
\begin{tabular}{c|l|l}
    \hline
    \textit{Region number in Power atlas} & \textit{Functional module} & \textit{Brain location}  \\
    \hline
    35 & Motor & Left cerebrum, frontal lobe, precentral gyrus \\
    151 & Visual & Left cerebrum, occipital lobe, lingual gyrus \\
    167 & Visual & Left cerebrum, occipital lobe, cuneus \\
    230 & Sub-cortical & Right cerebrum, sub-lobar, lentiform nucleus \\
    154 & Visual & Left cerebrum, occipital lobe, middle occipital gyrus \\
    38 & Motor & Left cerebrum, parietal lobe, postcentral gyrus \\
    79 & Default mode & Left cerebrum, temporal lobe, middle temporal gyrus \\
    58 & Cingulo-opercula & Left cerebrum, temporal lobe, superior temporal gyrus \\
    137 & Default mode & Left cerebrum, frontal lobe, inferior frontal gyrus \\
    175 & Fronto-parietal & Right cerebrum, frontal lobe, middle frontal gyrus \\
    96 & Default mode & Right cerebrum, parietal lobe, angular gyrus \\
    239 & Ventral attention & Right cerebrum, temporal lobe, middle temporal gyrus \\
    165 & Visual & Right cerebrum, occipital lobe, lingual gyrus \\
    52 & Cingulo-opercula & Right cerebrum, sub-lobar, insula \\
    126 & Default mode & Left cerebrum, temporal lobe, fusiform gyrus \\
    \hline
\end{tabular}
}
\caption{\textbf{Top panel}: GTP study analysis results for predictive mean squared error (MSE), coverage and (interval) width from 100 random splits. Noted that our approaches, ls-GPR1 and ls-GPR2, achieve significant reduction in predictive MSE at 0.05 or 0.10 level compared to competing methods. Our ls-GPR2 has similar coverage with the mf-GPR method but narrower predictive interval, indicating its greater confidence in the predicted values. The edge-GPR and the pca-GPR methods always have worse performance. 
\textbf{Bottom panel}: Supplementary analysis results on important regions in predicting resilience.
}
\label{tab:CDRISC}
\end{table}

\begin{figure}
    \centering
    \includegraphics[width=0.8\textwidth]{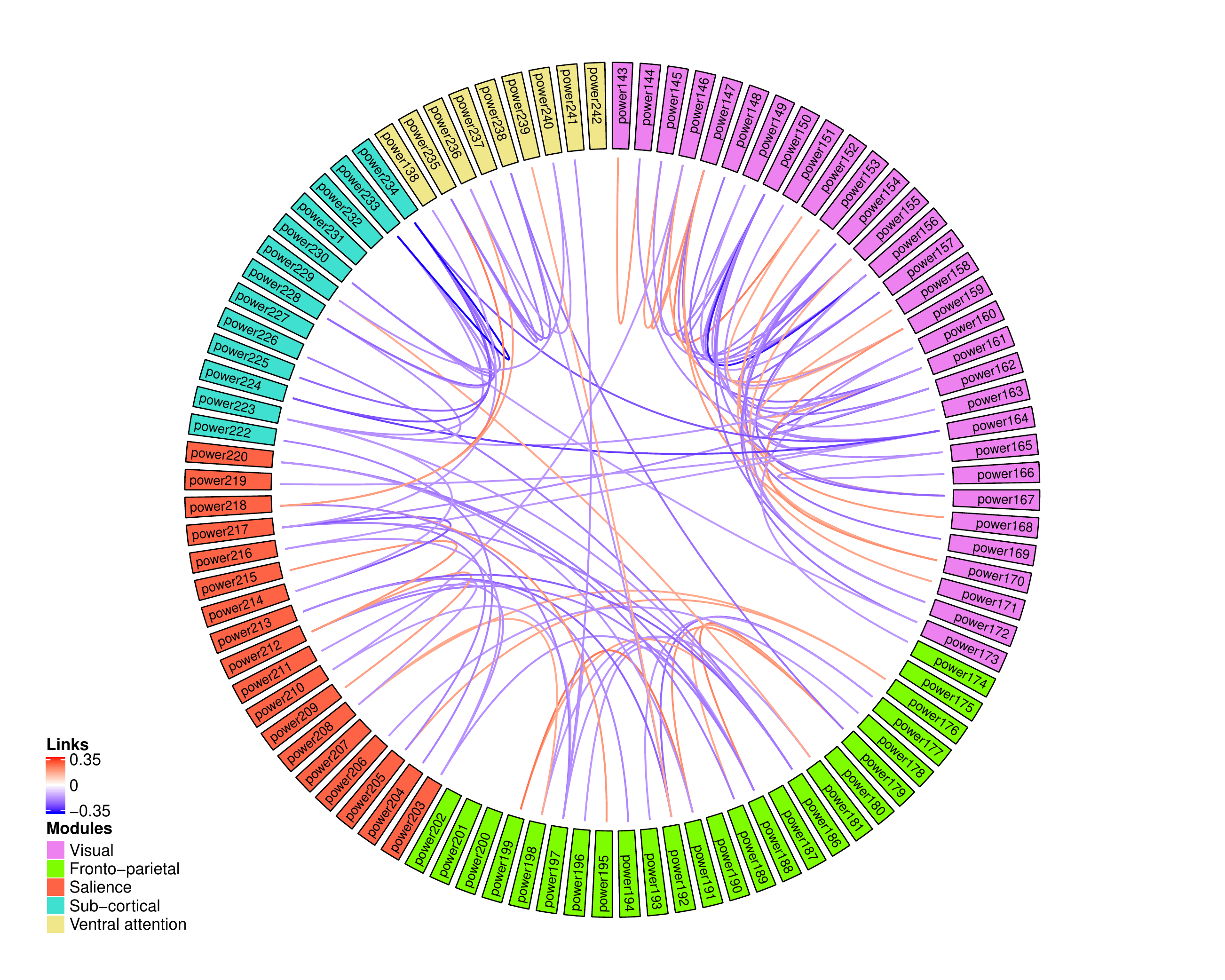}
    \caption{Circular plot for differences in functional connectivity between most and least resilient participants. Most differences for the visual network occur within-module, but still a few connected to salience and sub-cortical networks. The other modules see both within- and across-module differences.}
    \label{fig:chord}
\end{figure}

The right panels in Figure \ref{fig:abscor} display the difference in the estimated edge probabilities for participants with the highest and lowest resilience scores. In addition, Figure \ref{fig:chord} highlight the largest edge differences from selected functional modules. Most of the differences seem to be concentrated in the visual, fronto-parietal, salience, sub-cortical and ventral attention networks. Several important regions within these modules are also identified in terms of driving resilience via our supplementary analysis using modified latent scales described in Section \ref{subs:GTP}. Details can be found in the bottom panel of Table \ref{tab:CDRISC}. These findings are consistent with the literature on resilience and functional connectivity \citep{whalley2013fmri,cisler2013differential,van2013resilience}.

\section{Discussion}
We develop a novel two-stage approach for brain network covariate in prediction of a clinical phenotype. We are able to admit other covariates in the regression model and account for their non-linear interaction with the network-valued covariate via flexible Gaussian process kernel. Our simulation studies and real data example showed improved prediction performance over other existing methods. The supplementary analysis for the GTP data in identifying important brain regions also showed the advantage of having a node-specific representation of the networks.

We note that it is possible that the two-step approach could lead to error propagation. However, our primary goal is to regress the clinical outcome on lower dimensional representations of the brain network, and some inadequacies in the estimation of the latent scales are tolerated as long as it does not lead to significant decrease in prediction performance in the second stage. One can evaluate the propagated error by implementing a Gibbs sampler algorithm for the first stage model to obtain a set of posterior samples for the latent scales, which can then be treated as additional testing samples for the second stage model to compute the corresponding predictions. The variation in these predictions can be viewed as proxy propagated error.

For our future work, we can consider incorporating covariate information into estimation of the latent scales as inspired by \cite{wang2017bayesian}. Our first stage model easily permits adaption to changes in certain covariates in the prior structure. It needs not necessarily be the same set of covariates as in the second stage regression model since they carry different information in the procedure.

\section*{\normalsize{Appendix A. Data Augmented EM Algorithm for First Stage Model}}
\label{sec:appdA}
Without loss of generality, we omit the subject-specific subscripts in the first stage model, since the latent scales are derived independently for each subject. We assign a non-informative prior to the global intercept $a$, i.e. $\pi(a) \propto 1$. Independent normal priors are assigned to the latent scales, i.e. $u_{kl} \sim \mathrm{N}(0,\sigma_u^2), k<l,k,l=1,\ldots,p$. In our estimation, we use $\sigma_u^2=0.2$ that works well for a variety of applications, but one can also specify a hyperprior to estimate this parameter in a data adaptive manner. The likelihood can be expressed as:
\begin{equation}
\label{eqt:full_lik}
\pi(\boldsymbol e|a,U) = \prod_{k<l,k,l=1}^p \frac{\Big[exp\big(a+\boldsymbol{u}_k\boldsymbol{u}_l^T\big)\Big]^{e_{kl}}}{\Big[1+exp\big(a+\boldsymbol{u}_k\boldsymbol{u}_l^T\big)\Big]}, k<l, k,l=1,\ldots,p.
\end{equation}
Using Theorem 1 in \citet{polson2013bayesian}, we can introduce edge-specific latent P{\'o}lya-Gamma ($\mathrm{PG}(0,1)$) variables $\boldsymbol{\omega} = \{\omega_{k,l}: k<l\}$ and rewrite the augmented likelihood as:
\begin{equation}
\label{eqn:aug_lik}
\pi(\boldsymbol{e}|a,U,\boldsymbol{\omega}) = \prod_{k<l,k,l=1}^p 0.5\exp\bigg\{(e_{kl}-0.5)\big(a+\boldsymbol{u}_k\boldsymbol{u}_l^T\big) - 0.5\omega_{k,l}\big(a+\boldsymbol{u}_k\boldsymbol{u}_l^T\big)^2\bigg\}
\end{equation}
Based on this expression, we can develop an EM algorithm to obtain the MAP (maximum a posteriori probability) estimates for $a$ and $U$ via the M-step, while treating the latent P{\'o}lya-Gamma variables as missing variables that are approximated via the E-step. The $q$-th iteration is described below.

{\noindent \bf \uline{E step:}} We calculate the conditional expectation of the P{\'o}lya--Gamma variables as
\begin{equation}
\omega_{k,l}^{(q)} = \frac{1}{2\delta_{k,l}^{(q)}}\bigg[\frac{e^{\delta_{k,l}^{(q)}}-1}{e^{\delta_{k,l}^{(q)}}+1}\bigg], \mbox{ }\delta_{k,l}^{(q)} = a^{(q-1)}+\boldsymbol{u}_k^{(q-1)}\boldsymbol{u}_l^{(q-1)T}, k<l, k,l=1,\ldots,p.
\label{eqt:Estep}
\end{equation}

{\noindent \bf \uline{M step:}} We plug in $\{\omega_{k,l}^{(q)}: k<l\}$ to find the values of $a$ and $U$ that maximize the objective function. We first find the estimate for $a$ as:
\begin{equation}
a^{(q)} = \frac{\sum_{k<l,k,l=1}^p \Big[e_{kl}-0.5-\omega_{k,l}^{(q)}\boldsymbol{u}_k^{(q-1)}\boldsymbol{u}_l^{(q-1)T}\Big]}{\sum_{k<l,k,l=1}^p \omega_{k,l}^{(q)}}, k<l, k,l=1,\ldots,p.
\label{eqt:Mstep_a}
\end{equation}

As noted earlier, the first element of every latent scale is fixed at a certain pre-specified value $b$ and does not need to be updated. Thus we denote latent scale for the $k$th node omitting the first element as $\boldsymbol{u}_{k(-1)}$. The latent scales are updated iteratively from first node to the last one as:
\begin{equation}
\begin{split}
\boldsymbol{u}_{k(-1)}^{(q)} = &\bigg\{\sum_{j<k} \Big[e_{jk}-0.5-(a^{(q)}+b^2)\omega_{j,k}^{(q)}\Big]\boldsymbol{u}_{j(-1)}^{(q)}+\sum_{j>k} \Big[e_{jk}-0.5-(a^{(q)}+b^2)\omega_{j,k}^{(q)}\Big]\boldsymbol{u}_{j(-1)}^{(q-1)}\bigg\}\\
&\bigg\{\sum_{j<k} \Big[\omega_{j,k}^{(q)}\boldsymbol{u}_{j(-1)}^{(q)T}\boldsymbol{u}_{j(-1)}^{(q)} + \sigma_u^{-2}\mathrm{I}_{(d-1)}   \Big]+\sum_{j>k} \Big[\omega_{j,k}^{(q)}\boldsymbol{u}_{j(-1)}^{(q-1)T}\boldsymbol{u}_{j(-1)}^{(q-1)} + \sigma_u^{-2}\mathrm{I}_{(d-1)}   \Big]\bigg\}^{-1}
\label{eqt:Mstep_U}
\end{split}
\end{equation}

We use the multidimensional scaling (MDS) method \citep{mardia1978some} to find the initial positions of the latent scales in $\Re^d$ such that the distance matrix is best preserved. The distance matrix between nodes are calculated as the shortest undirected paths based on the binary network $G$. We denote the estimates from the algorithm as $\hat{a}$ and $\hat{U}$.

\section*{\normalsize{Appendix B. Gradient Descent Algorithm with Armijo Line Search for Second Stage MLE}}
\label{sec:appdB}
The objective function is in fact the negative log-likelihood, which is of the form:
\begin{equation}
\begin{split}
    O(\boldsymbol{\theta}) &= 0.5log\big|K(\boldsymbol{\theta})\big| + \frac{1}{2}\boldsymbol{y}^T K(\boldsymbol{\theta})^{-1}\boldsymbol{y}\\
    K(\boldsymbol{\theta}) &= e^{-\theta_1}\mathrm{I}_n + exp\Big[\theta_2 - e^{\theta_3}E_u - e^{\theta_4}E_a - e^{\theta_5}E_z\Big]\\
    E_u(i,i^\prime) = \big|\big|\hat{U}_i - \hat{U}_{i^\prime}\big|\big|_F^2,\mbox{ } &E_a(i,i^\prime) = (\hat{a}_i - \hat{a}_{i^\prime})^2,\mbox{ }  E_z(i,i^\prime) = \big|\big|\boldsymbol{z}_i - \boldsymbol{z}_{i^\prime}\big|\big|_2^2,
    \mbox{ } i, i^\prime = 1,\cdots,n,
\end{split}
\label{eqt:obj_func}
\end{equation}
where $\boldsymbol{\theta} = (\theta_1, \theta_2, \theta_3, \theta_4, \theta_5) = (log\tau,log\psi_1,log\psi_u,log\psi_a,log\psi_z)$, and 
$|\cdot|$ denotes the matrix determinant. The partial derivatives of the objective function with respect to the parameters are used for deriving the MLE and they have the following forms
\begin{equation}
\frac{\partial{O(\boldsymbol{\theta})}}{\partial{\theta_j}} = 0.5\rmn{trace}\bigg[(K(\boldsymbol{\theta})^{-1} - \boldsymbol{\alpha}\boldsymbol{\alpha}^T)\frac{\partial{K(\boldsymbol{\theta})}}{\partial{\theta_j}}\bigg],\mbox{ } \boldsymbol{\alpha}=K(\boldsymbol{\theta})^{-1}\boldsymbol{y}, j=1,\ldots,5,
\label{eqt:grd_form}
\end{equation}
where $\frac{\partial{f}}{\partial{x}}$ denotes the partial derivative of $f$ with respect to $x$. More specifically
\begin{equation}
\begin{split}
    \frac{\partial{K(\boldsymbol{\theta})}}{\partial{\theta_1}} &= -e^{-\theta_1}\mathrm{I}_n,\mbox{ } \frac{\partial{K(\boldsymbol{\theta})}}{\partial{\theta_2}} = e^E,\mbox{ } \frac{\partial{K(\boldsymbol{\theta})}}{\partial{\theta_3}} = e^E \odot (-e^{\theta_3}E_u)\\
    \frac{\partial{K(\boldsymbol{\theta})}}{\partial{\theta_4}} &= e^E\odot (-e^{\theta_4}E_a),\mbox{ } \frac{\partial{K(\boldsymbol{\theta})}}{\partial{\theta_5}} = e^E\odot (-e^{\theta_5}E_z)\\
    E &= \theta_2 - e^{\theta_3}E_u - e^{\theta_4}E_a - e^{\theta_5}E_z,
\end{split}
\label{eqt:grd_deriv}
\end{equation}
where $\odot$ denotes the Hadamard product. Algorithm \ref{alg:mle} minimized the objective function $O(\boldsymbol{\theta})$ to find the MLE for the log-scaled parameters. We then transformed back by taking exponential to obtain the MLE of the original parameters.
\begin{algorithm}
	\caption{Gradient descent with Armijo line search}
	\label{alg:mle}
	\begin{spacing}{1}
	\begin{algorithmic}
		\STATE \textbf{Input:} vector $\boldsymbol{\theta},\mbox{ } maxiter=1000,\mbox{ } \nu=10^{-8},\mbox{ } lr=1$
		\FOR{$g=1,\ldots,maxiter$}
		   \STATE $f = O(\boldsymbol{\theta}),\mbox{ } \boldsymbol{df} = \frac{\partial{O(\boldsymbol{\theta})}}{\partial{\boldsymbol{\theta}}}$
           \IF{$O(\boldsymbol{\theta}-lr\cdot \boldsymbol{df}) < f$}
              \STATE $\boldsymbol{\theta} = \boldsymbol{\theta} - lr\cdot \boldsymbol{df}$
              \STATE $lr = 1.5lr$
              \IF{$|f - O(\boldsymbol{\theta})| < \nu$} \STATE \textbf{break} \ENDIF
              \STATE \textbf{next}
           \ENDIF
           \WHILE{$O(\boldsymbol{\theta}-lr\cdot \boldsymbol{df}) \geq f$} \STATE $lr = 0.5lr$ \ENDWHILE
           \STATE $\boldsymbol{\theta} = \boldsymbol{\theta} - lr\cdot \boldsymbol{df}$
           \IF{$|f - O(\boldsymbol{\theta})| < \nu$} \STATE \textbf{break} \ENDIF
        \ENDFOR
		\STATE \textbf{Output:} vector $\boldsymbol{\theta}$
	\end{algorithmic}
	\end{spacing}
\end{algorithm}

\section*{\normalsize{Appendix C. Gibbs Sampler for Second Stage Parameters}}
\label{sec:appdC}
We denote the Gaussian process atoms as $\boldsymbol{\phi} = \Big(\phi(\hat{a}_1, \hat{U}_1, \boldsymbol z_1),\cdots,\phi(\hat{a}_n, \hat{U}_n, \boldsymbol z_n)\Big)^T$. The algorithm iterates between the following steps:
\begin{enumerate}
    \item Update the noise precision $\tau$ from Gamma distribution with parameters ($a_\tau + 0.5n$) and ($b_\tau + 0.5\big|\big|\boldsymbol{y}-\boldsymbol{\phi}\big|\big|_2^2$).
    \item Update the global scale parameter $\psi_1$ from inverse Gamma distribution with parameters ($a_{\psi_1} + 0.5n$) and ($b_{\psi_1} + 0.5\boldsymbol{\phi}^T E_0^{-1}\boldsymbol{\phi}$) where $E_0 = exp\Big(-\psi_u E_u - \psi_a E_a - \psi_z E_z\Big)$. $E_u, E_a, E_z$ are defined in Equation (\ref{eqt:obj_func}).
    \item Draw a candidate $\psi_u^*$ where $log\psi_u^* \sim \mathrm{N}(log\psi_u, 0.01^2)$. Accept the candidate with probability
    \begin{equation*}
        min\Bigg(1, \frac{\big|\psi_1E_0^*+\tau^{-1}\rmn{I}_n\big|^{-0.5}exp\{-0.5\boldsymbol{y}^T (\psi_1E_0^*+\tau^{-1}\rmn{I}_n)^{-1}\boldsymbol {y}\}}{\big|\psi1E_0+\tau^{-1}\rmn{I}_n\big|^{-0.5}exp\{-0.5\boldsymbol{y}^T (\psi1E_0+\tau^{-1}\rmn{I}_n)^{-1}\boldsymbol{y}\}}\Bigg)
    \end{equation*}
    where $E_0^* = exp\Big(-\psi_u^* E_u - \psi_a E_a - \psi_z E_z\Big)$. Same procedure for updating $\psi_a$ and $\psi_z$.
    \item Update the Gaussian process atoms $\boldsymbol{\phi}$ from multivariate normal distribution with mean $\Big[\tau^{-1}\psi_1^{-1}E_0^{-1}+\mathrm{I}_n\Big]^{-1}\boldsymbol{y}$ and covariance $\Big[\psi_1^{-1}E_0^{-1}+\tau\mathrm{I}_n\Big]^{-1}$.
\end{enumerate}
For our supplementary analysis using the modified latent scales as described in Section \ref{subs:GTP}, we need two more steps in the Gibbs Sampler to update the node inclusion indicators $\beta_{i1},\ldots,\beta_{ip}$ and the hyperparameter $\pi^*$ as illustrated below:
\begin{enumerate}
    \setcounter{enumi}{4}
    \item While keeping all other indicators fixed, update $\beta_j$ from Bernoulli distribution with probability
    \begin{equation*}
         min\Bigg(1,\mbox{ } \frac{\pi^*\rmn{lik}_{j1}}{\pi^*\rmn{lik}_{j1}+(1-\pi^*)\rmn{lik}_{j0}}\Bigg)
    \end{equation*}
    where $$\rmn{lik}_{j1} = \big|\psi_1E_0^{(j1)}+\tau^{-1}\rmn{I}_n\big|^{-0.5}exp\{-0.5\boldsymbol{y}^T (\psi_1E_0^{(j1)}+\tau^{-1}\rmn{I}_n)^{-1}\boldsymbol {y}\},$$
    $$\rmn{lik}_{j0} = \big|\psi1E_0^{(j0)}+\tau^{-1}\rmn{I}_n\big|^{-0.5}exp\{-0.5\boldsymbol{y}^T (\psi1E_0^{(j0)}+\tau^{-1}\rmn{I}_n)^{-1}\boldsymbol{y}\}.$$ In addition, we have $E_0^{(j1)}=exp\Big(-\psi_u E_u^{(j1)} - \psi_a E_a - \psi_z E_z\Big)$ and $E_0^{(j0)}=exp\Big(-\psi_u E_u^{(j0)} - \psi_a E_a - \psi_z E_z\Big)$. Here $E_u^{(j1)}$ is calculated as in Equation \ref{eqt:obj_func} but with modified latent scales $U_i\boldsymbol{\beta}^{(j1)} = (\beta_1\boldsymbol{u}_{i1}^T,\ldots,\boldsymbol{u}_{ij}^T,\ldots,\beta_p\boldsymbol{u}_{ip}^T)^T$, and $E_u^{(j0)}$ is calculated with modified latent scales whose $j$-th row is replaced with a $\mathbf{0}$ vector.
    \item Update the hyperparameter $\pi^*$ from Beta distribution with parameters $(a_\pi+\sum_{j=1}^p\beta_j)$ and $(b_\pi+p-\sum_{j=1}^p\beta_j)$.
\end{enumerate}

\section*{\normalsize{Appendix D. Prediction for Testing Samples}}
\label{sec:appdD}
The prediction for additional subjects involves two steps. First step takes the upper triangular vector of binary network matrices $\boldsymbol{e}_1^*, \cdots, \boldsymbol{e}_m^*$ as input of first stage model and obtain the estimates of intercept and latent scales. These estimates are then used in the second step, together with other supplementary covariates, as input of second stage model and eventually obtain the predicted values of the response variable. From the property of Gaussian process, the response of the additional subjects $\boldsymbol{y}^* = (y_1^*,\cdots,y_m^*)^T$ and the response of the original $n$ subjects $\boldsymbol{y}$ jointly follow a multivariate normal distribution as
\begin{equation}
\begin{split}
    \begin{pmatrix}\boldsymbol{y}\\ \boldsymbol{y}^*\end{pmatrix} &\sim \mathrm{N}\vast(\mathbf{0}_{(n+m)},\begin{pmatrix} K+\tau^{-1}\mathrm{I}_n & K_*^T \\ K_* & K_{**}+\tau^{-1}\mathrm{I}_n \end{pmatrix} \vast),\\
    K_*(q,i) &= \psi_1 exp\Big\{-\psi_u \big|\big|U_q^* - U_i\big|\big|_F^2 - \psi_a (a_q^* - a_i)^2 - \psi_z \big|\big|\boldsymbol z_q^* - \boldsymbol z_i\big|\big|_2^2\Big\},\\
    K_{**}(q,q^\prime) &= \psi_1 exp\Big\{-\psi_u \big|\big|U_q^* - U_{q^\prime}^*\big|\big|_F^2 - \psi_a (a_q^* - a_{q^\prime}^*)^2 - \psi_z \big|\big|\boldsymbol z_q^* - \boldsymbol z_{q^\prime}^*\big|\big|_2^2\Big\},\\
    &\quad \quad i=1,\cdots,n,\mbox{ } q,q^\prime = 1,\cdots,m
\end{split}
\label{eqt:joint_norm}
\end{equation}
Therefore, the conditional mean can be expressed as:
\begin{equation}
    \mathbf{E}(\boldsymbol{y}^*|\boldsymbol{y}) = K_*(K+\tau^{-1}\mathrm{I}_n)^{-1}\boldsymbol{y}
\label{eqt:norm_mean}
\end{equation}
The expression above is based on the marginal distribution of the outcome. We use this expression with the parameters obtained from MLE. We can also base the prediction expression on the Gaussian process atoms $\boldsymbol{\phi}$. Then the computation would be
\begin{equation}
    \begin{pmatrix}\boldsymbol{\phi}\\ \boldsymbol{y}^*\end{pmatrix} \sim \mathrm{N}\vast(\mathbf{0}_{(n+m)},\begin{pmatrix} K & K_*^T \\ K_* & K_{**}+\tau^{-1}\mathrm{I}_n \end{pmatrix} \vast),\mbox{ } \mathbf{E}(\boldsymbol{y}^*|\boldsymbol{\phi}) = K_*K^{-1}\boldsymbol{\phi}
\label{eqt:joint_norm2}
\end{equation}
This expression can be used for in-sample prediction with the Gibbs sampler algorithm.

\bibliographystyle{apalike}

\end{document}